\documentclass{iau_FM}
\usepackage[utf8]{inputenc}
\usepackage{graphicx}
\usepackage{amsmath}
\usepackage{amssymb}

\title[Cosmic-ray propagation in the turbulent intergalactic medium] 
{Cosmic-ray propagation in the turbulent intergalactic medium}

\author[R. Alves~Batista {\it et al.}] 
{Rafael Alves~Batista$^{1,*}$,
Elisabete M. de~Gouveia~Dal~Pino$^{1,\dagger}$,
Klaus Dolag$^{2,3}$,
Saqib Hussain$^1$}

\affiliation{$^{1}$Universidade de São Paulo - Instituto de Astronomia, Geofísica e Ciências Atmosféricas; Rua do Matão, 1226, 05508-090, São Paulo-SP, Brazil  \\[\affilskip]
$^2$ Max Planck Institute for Astrophysics, Karl-Schwarzschild-Str 1, 85741 Garching, Germany \\[\affilskip]
$^3$ Universitäts-Sternwarte München, Scheinerstraße 1, 81679, München, Germany \\[\affilskip]
$^*$ email: {\tt rafael.ab@usp.br}\\
$^\dagger$ email: {\tt dalpino@iag.usp.br}
}

\pubyear{2015}
\jname{Astronomy in Focus, Volume 1} 
\editors{Piero Benvenuti, ed.}
\begin{document}

\maketitle

\begin{abstract}

Cosmic rays (CRs) may be used to infer properties of intervening cosmic magnetic fields. Conversely, understanding the effects of magnetic fields on the propagation of high-energy CRs is crucial to elucidate their origin. In the present work we investigate the role of intracluster magnetic fields on the propagation of CRs with energies between $10^{16}$ and $10^{18.5}$ eV. We look for possible signatures of a transition in the CR propagation regime, from diffusive to ballistic. Finally, we discuss the consequences of the confinement of high-energy CRs in clusters and superclusters for the production of gamma rays and neutrinos.

\keywords{magnetic fields,  large-scale structure of universe, MHD, galaxies: clusters: general}
\end{abstract}

\firstsection 
\section{Introduction}

Cosmic magnetic fields are ubiquitous in the universe and their intensities span from  $10^{-10} \; \text{G}$ in  extended filaments, which are possibly the largest magnetised structures in the universe, up to around $10^{15} \; \text{G}$ in magnetars. Tiny magnetic fields ($\sim 10^{-17} \; \text{G}$) can also exist in the vast regions devoided of matter, the so-called cosmic voids (see \cite{barai2018a} and references therein). Presently little is known about the origin and evolution of these diffuse intergalactic magnetic fields (IGMFs). 

CRs are produced via shock and turbulent acceleration processes in  active galaxies and even in the more diffuse regions of the intergalactic medium (IGM) like relics and haloes (e.g.~\cite{brueggen2015a}). They are deflected by the pervasive magnetic fields and thus can be used to constrain their strength and coherence lengths.    
 
Strong evidence exists supporting an extragalactic origin of CRs with energies $E \gtrsim 8 \times 10^{18} \; \text{eV}$~(\cite{auger2017a}), but it is not clear yet at which energies there is a transition between galactic and extragalactic CRs. Moreover, there are indications of a possible third component comprised of nearby extragalactic sources potentially responsible for the CR spectrum at energies between the second knee ($E \sim 10^{17} \; \text{eV}$) and the ankle ($E \sim 10^{18.5} \; \text{eV}$)~(\cite{deligny2014a}). At energies below $E \sim Z \times 10^{17} \; \text{eV}$, wherein $Z$ is the atomic number of the cosmic-ray nucleus, it is widely believed that galactic sources dominate the spectrum as their Larmor radii are about the size of our galaxy. 

A number of authors have performed magnetohydrodynamical (MHD) simulations of large-scale structure formation which were subsequently used for the propagation of CRs, particularly the highest-energy ones (UHECRs). For instance,~\cite{dolag2004a} and~\cite{sigl2003a} have reached diverging conclusions regarding the effects of extragalactic magnetic fields on UHECR propagation. The former concluded that deflections should be small in most of the sky (see also \cite{medinatanco98a}), whereas the latter concluded that they are large, thereby disfavouring the identification of individual sources of UHECRs. In more recent work, \cite{hackstein2016a} obtained results that favour small deflections, which is consistent with their later work using constrained MHD simulations~(\cite{hackstein2017a}). A more pessimistic scenario has been studied by~\cite{alvesbatista2017a}, who concluded that even if the magnetic fields in cosmic voids, which fill most of the volume of the universe, are high, the deflection of UHE protons with $E \gtrsim 5 \times 10^{19} \; \text{eV}$ would be $\lesssim 15^\circ$ in about 10-50\% of the sky, depending on the magnetic power spectrum. This would enable the identification of UHECR sources even in the most pessimistic scenarios.

The distribution of extragalactic magnetic fields is largely uncertain and many differences exist across models. 
A comprehensive review on the topic was presented by~\cite{vazza2017a}, and a detailed discussion of these uncertainties on UHECR propagation was given by~\cite{alvesbatista2017a}. 
Other properties of the magnetised IGM may also affect the propagation of CRs. This includes magnetic helicity, related to the topology of the magnetic field lines, which can leave an imprint in the large-scale distribution of CR arrival directions for some specific source distributions~(\cite{alvesbatista2018a}), as well as plasma instabilities (\cite{brunetti2015a}). 

It is not clear at which energy the transition between the ballistic and diffusive regimes of CR propagation occur. Also, there is an energy below which the flux of extragalactic CRs is strongly suppressed due to magnetic horizon effects, but this depends on the strength and coherence length of the magnetic fields, as well as the distribution of sources. \cite{alvesbatista2014a} have shown that for sources distributed approximately uniformly, this energy is $E \lesssim 10^{18} \; \text{eV}$ for common extragalactic magnetic field models obtained via MHD simulations. Note, however, that this result depends on the distribution of magnetic fields between the closest sources and Earth, so that magnetic horizon effects may play a role for highly inhomogeneous distributions of sources.

\section{Setup of the simulations}

Using 3D cosmological MHD simulations done by~\cite{dolag2004a}, obtained with GADGET~(\cite{springel2005a}), we investigate the propagation of CRs with energies $10^{16} \lesssim E / \text{eV} \lesssim 10^{18.5}$, whose origins are unknown and about which few studies have been done. 

We perform a numerical study of the propagation of CRs in aforementioned MHD background. We employ the CRPropa code~(\cite{alvesbatista2016a}). We consider adiabatic energy losses due to the expansion of the universe. Photohadronic and photonuclear interactions between high-energy CRs and photons from the CMB, EBL, and from the clusters are neglected in this first study. Interactions between CMB/EBL photons with CRs are virtually negligible at these energies. 

The cosmological simulations we use here have been previously used by~\cite{dolag2004a} for UHECR propagation studies. They correspond to a volume of $\sim (140 \; \text{Mpc})^3$. The initial conditions of the simulation are such that the relative positions between the main structures and Earth are roughly preserved. We select a sample of clusters from the simulation for the analysis, and choose the Virgo cluster to discuss the results.

\section{Results}

\begin{figure}[htb]
  \centering
  \includegraphics[width=0.32\columnwidth]{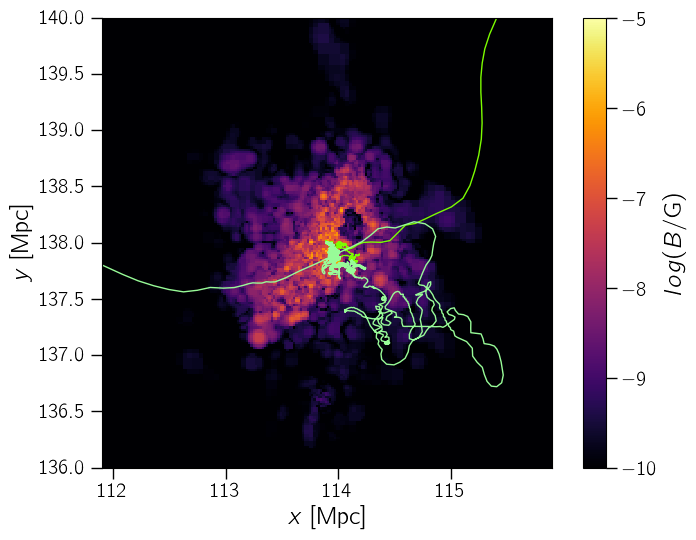}
  \includegraphics[width=0.32\columnwidth]{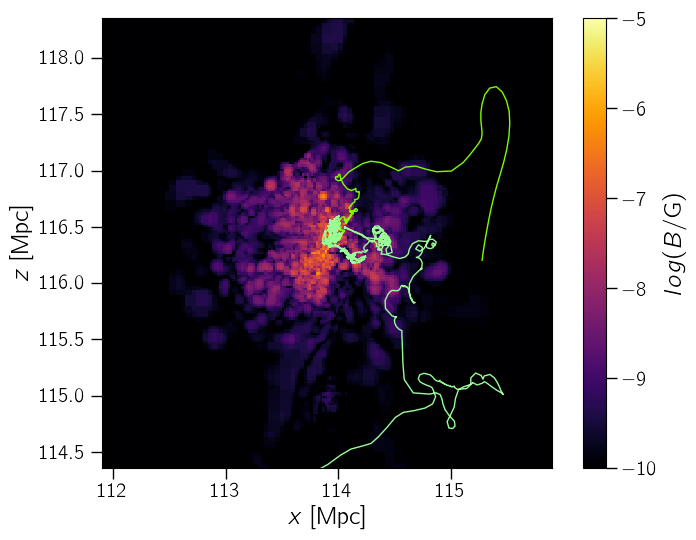}
  \includegraphics[width=0.32\columnwidth]{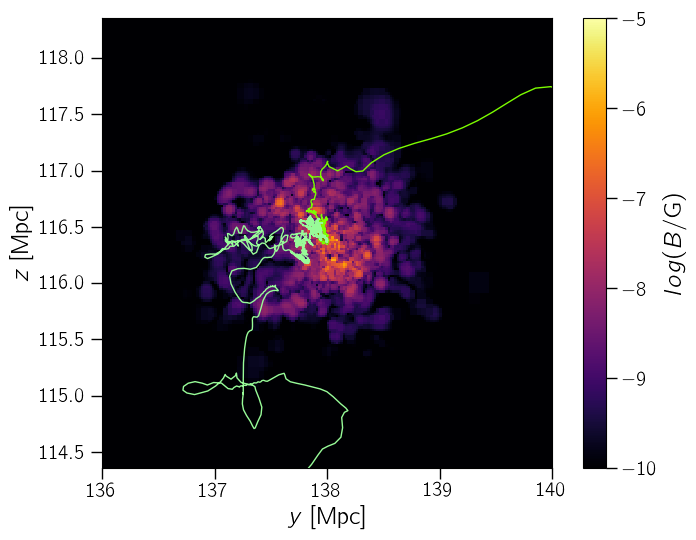}
  \includegraphics[width=0.32\columnwidth]{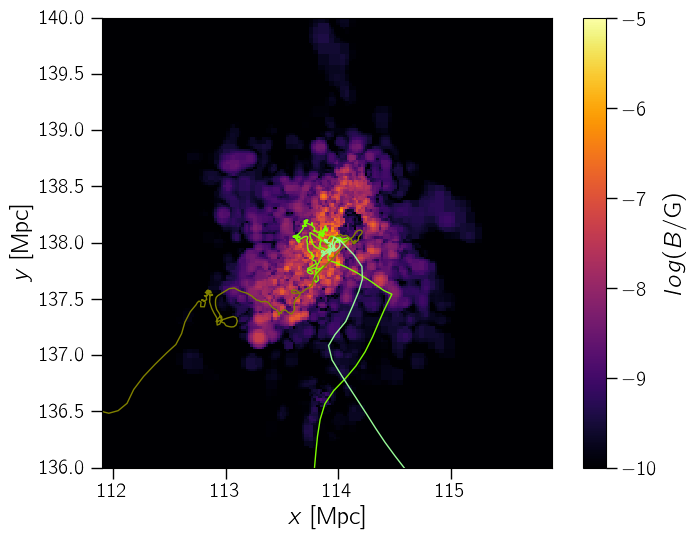}
  \includegraphics[width=0.32\columnwidth]{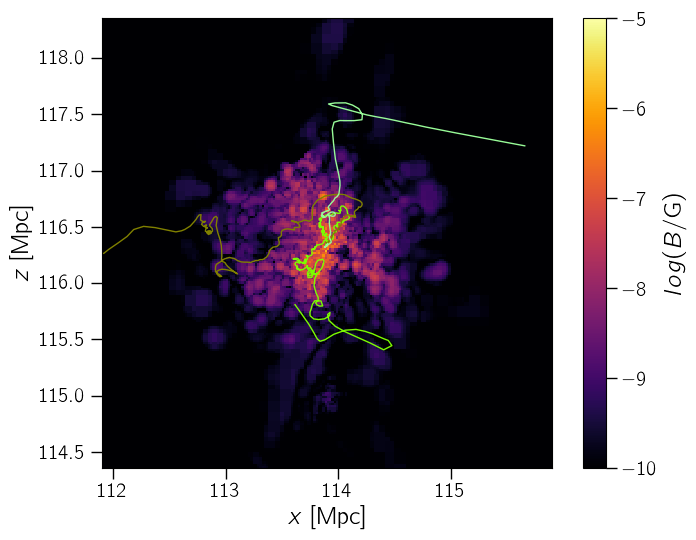}
  \includegraphics[width=0.32\columnwidth]{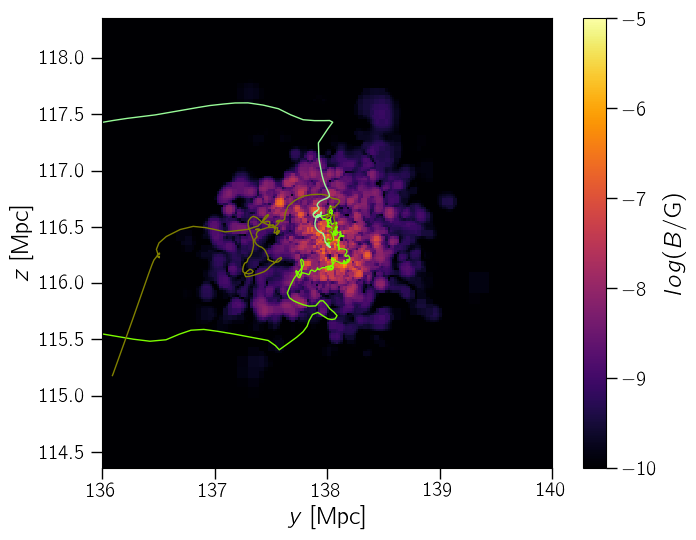}
  \caption{Slices $xy$ (left), $xz$ (middle) e $yz$ (right column) of the simulated volume. The colour scale corresponds to the intensity of the field in these regions. Trajectories of three CR protons are represented through the green lines with different shades. The scenario in the upper row corresponds to $E=10^{17} \; \text{eV}$, whereas the lower row is for $E=10^{18} \; \text{eV}$.}
  \label{fig:traj}
\end{figure}

In Fig.~\ref{fig:traj} the trajectories of three cosmic-ray protons are shown for $E = 10^{17} \; \text{eV}$ and $E = 10^{18} \; \text{eV}$. One can see in Fig.~\ref{fig:traj} that the propagation is approximately diffusive near the source, where the intensity of the fields is higher, transitioning towards a rectlinear propagation as energy increases. 

\begin{figure}[htb]
  \centering
  \includegraphics[width=0.45\columnwidth]{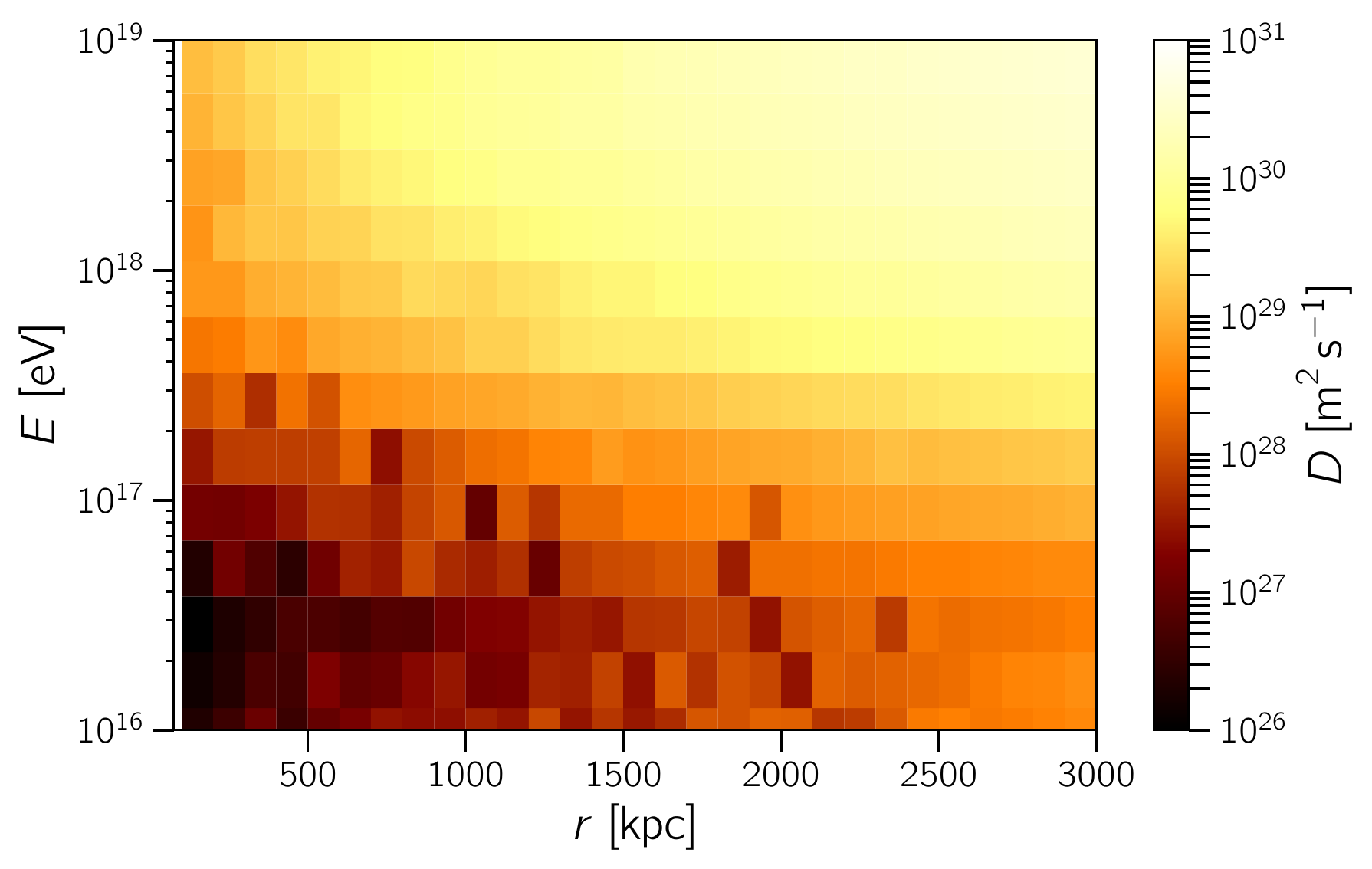}
  \caption{Diffusion coefficient as a function of the cosmic-ray energy and distance to the centre of the cluster for the case of protons with an $E^{-1}$ spectrum leaving the Virgo cluster.}
  \label{fig:diffusion}
\end{figure}

The diffusion coefficient is energy-dependent and combines two regimes, quasi-linear ($D \propto E^{1/3}$), and non-resonant ($D \propto E^2$). The diffusion coefficient can be written as $D = \langle r^2 \rangle / 6t$, wherein $r$ is the displacement of the CR with respect to its initial position, and $t$ is the time it takes to move a distance $r$. CRs can escape the cluster if $t$ is less than the age of the cluster, assumed here to be of the order of a Hubble time; $r \simeq 1 \; \text{Mpc}$, the typical size of a cluster. This implies that for $D \lesssim 10^{27} \; \text{m}^2 \, \text{s}^{-1}$ the diffusion time is comparable to one Hubble time. The behaviour of the diffusion coefficients for different combinations of energy and distance to the centre of the cluster is shown in Fig.~\ref{fig:diffusion}. 

Note that in the central regions of the cluster ($r \lesssim 500$ \; \text{kpc}) the CRs tend to be confined longer compared to those in the cluster outskirts. The energy at which CRs cease to be magnetically confined is $E \sim 10^{17} \; \text{eV}$. 
While this result is somewhat trivial, in the sense that it follows from considerations of the strength of magnetic fields and the size of clusters, it enables us to reach important conclusions. First, all CRs with energies $E \lesssim 10^{17} \; \text{eV}$ should originate within our local supercluster. Second, if our supercluster contains sources of high-energy CRs, the low-energy suppression of the extragalactic CR spectrum should not occur at energies much larger than $\sim 10^{17} \; \text{eV}$.

\section{Conclusions \& Outlook}

In the present work we have presented preliminary results of a first investigation of the effects of intracluster magnetic fields on CR propagation, using cosmological MHD simulations. We have estimated the diffusion coefficients of CRs in clusters. One of our main results is that below $E \sim 10^{17} \; \text{eV}$ CRs cannot escape the innermost regions of clusters. As a consequence, CRs with $E \lesssim 10^{17} \; \text{eV}$ originate either within the Milky Way or in our local cluster. We have also concluded that at these energies CRs are confined for a time comparable to the age of the cluster.  

Thermal UV and X-ray photons produced in the hot ICM may serve as target fields for CR interactions depending on their density; in the central regions of clusters, in particular, this effect may be considerable. 
Employing the numerical tools here described, we are currently performing a detailed analysis of high-energy CR propagation in galaxy clusters taking into account the photohadronic, photonuclear, and hadronuclear interactions between the CRs and the intracluster gas/photons, which may produce a substantial number of secondary high-energy particles including gamma rays and neutrinos.

\section*{Acknowledgements}

This work is supported by the São Paulo Research Foundation (FAPESP) grants \#2017/12828-4 and \#2013/10559-5, and by CNPq grants (\#306598/2009-4).

\end{document}